\documentclass{article}
\usepackage{latexsym}
\usepackage{amsmath,amsthm}
\usepackage{amssymb}
\usepackage{graphics}
\usepackage{color}
\newtheorem{theorem}{Theorem}[section]
\newtheorem{proposition}{Proposition}[section]

\newtheorem{lemma}{Lemma}[section]

\begin{document}
\title{Strong cosmic censorship for $T^2$-symmetric
cosmological spacetimes with
collisionless matter}
\author{Mihalis Dafermos\thanks{University of Cambridge,
Department of Pure Mathematics and Mathematical Statistics,
Wilberforce Road, Cambridge CB3 0WB United Kingdom} \and Alan D.~Rendall\thanks{
Max-Planck-Institut f\"ur Gravitationsphysik
(Albert-Einstein-Institut), 
Am M\"uhlenberg 1,
D-14476 Potsdam}}
\maketitle
\begin{abstract}
We prove strong cosmic censorship for
$T^2$-symmetric cosmological spacetimes (with spatial topology $T^3$
and vanishing cosmological constant $\Lambda$) with collisionless matter. Gowdy
symmetric spacetimes constitute a special case.
The formulation of the conjecture is in terms of generic $C^2$-inextendibility of the metric.
Our argument exploits 
a rigidity property of Cauchy horizons, inherited from the Killing fields. 
\end{abstract}

\section{Introduction}
Strong cosmic censorship is one of the fundamental open problems
of classical general relativity. Properly formulated~\cite{chr:givp}, it is
the conjecture that the maximal development of generic compact or asymptotically flat initial
data for suitable Einstein-matter systems
be inextendible as a suitably regular Lorentzian manifold.

In recent years, progress has been made when the initial data are restricted to 
symmetry classes, in particular, spherical~\cite{ns, md:si,md:cbh} and Gowdy~\cite{cim, hr} 
symmetry. The nature
of the difficulties in these two classes is very different. In addition to the case of a horizon
arising from a singular point on the  centre, Cauchy horizons in spherical symmetry
can arise on account of a global property of the causal geometry of the
Lorentzian quotient manifold of group orbits. An example is provided by
the Reissner-Nordstr\"om solution. The stability or instability of this phenomenon depends on 
what is essentially a completely
global analysis.\footnote{This already indicates that strong cosmic censorship in its
full generality
can never be approached by a local analysis in the style of
the so-called ``BKL proposal''.} In the Gowdy case with spatial topology $T^3$, Cauchy horizon formation is a local
phenomenon from the point of view of the quotient, and is related
to the group orbits becoming null. 
The remarkable recent progress~\cite{hr}
 on the $C^2$-inextendibility version of
  cosmic censorship in the Gowdy case for the vacuum equations
rests on a detailed asymptotic analysis of the local 
behaviour of a solution near points of the past boundary of
the maximal development. 

Gowdy symmetric spacetimes are a sub-class of so-called $T^2$-symmetric
spacetimes, i.e.~spacetimes which admit a torus action. 
The asymptotic analysis of~\cite{hr} seems prohibitively difficult in this more general case,
leaving to far in the future the task of pursuing this approach for proving strong cosmic censorship
in this class.

The aim of the present paper is to show that this difficult analysis can in fact be completely
circumvented if one is willing to couple the Einstein equations with the Vlasov
equation, i.e.~to consider spacetimes with collisionless matter.
We will thus here give an elementary proof of
strong cosmic censorship (in the $C^2$-inextendibility formulation)
for general $T^2$-symmetric spacetimes (with spatial 
topology $T^3$) solving
the Einstein-Vlasov system.
The proof relies on a previous characterization of the boundary of the maximal development
proven by Weaver~\cite{mw}.\footnote{Strictly 
speaking, the characterization of the boundary in~\cite{mw}
concerns a class of initial data too special, for in the present
paper it will be assumed that the support of $f$ on the mass shell is non-compact.
These results can be easily adapted, however. See for instance~\cite{dr3} for this
adaptation in the surface
symmetric case.}
Our method can be expected to apply when additional matter
fields are also coupled, but the presence of the Vlasov field is essential. 

The main idea of the method is quite simple: 
Let us suppose that our spacetime is $C^2$-past extendible with Cauchy horizon
$\mathcal{H}^-$. The characterization~\cite{mw} of the past
boundary of the maximal development allows us to deduce that  there is a null
vector in the span of the Killing fields for a dense open subset of $\mathcal{H}^-$.
It turns out that this fact gives the Cauchy horizon considerable rigidity, in particular,
${\rm Ric}(K,K)\le 0$ where $K$ denotes the null generator of $\mathcal{H}^-$ at regular points, 
and thus,
by the null convergence condition, ${\rm Ric}(K,K)=0$. On the other hand, we can bound the
Ricci curvature away from $0$ by following geodesics back to initial data, provided
that the matter is supported initially on a suitably large portion of the mass shell, specifically,
that its support intersects every open set. By exploiting conservation of the inner product
of velocity with a Killing vector along any geodesic, we may weaken this to the condition that the matter 
intersects every open set of sufficiently small tangential momentum. The
contradiction yields strong cosmic censorship.

The arguments of the present paper have been adapted from our recent work~\cite{dr3} on
strong cosmic censorship in surface symmetry, in particular, the case of hyperbolic
symmetry. In the context of the type of arguments employed here, the hyperbolic symmetric case
is in fact considerably more complicated than the $T^2$ symmetric case,
because its Lie algebra is non-abelian and Killing vectors can vanish. 
Moreover, the hyperbolic symmetric case also may admit Cauchy horizons on account of global phenomena, 
and these must be treated by a separate and very different method. The 
reader is strongly encouraged to look at~\cite{dr3}.

\section{The Einstein-Vlasov system}
Let $(\mathcal{M},g)$ be a $4$-dimensional spacetime with $C^2$ metric.
Let $P\subset T\mathcal{M}$ denote the set of all future directed
timelike vectors of length $-1$. We will call $P$ the \emph{mass shell}. 
Let $f$ denote a nonnegative function
$f:P\to{\mathbb R}$. 
We say that $\{(\mathcal{M},g),f\}$ satisfies the Einstein-Vlasov system (with 
vanishing cosmological constant $\Lambda$)
if
\begin{equation}
\label{Eeq}
R_{\mu\nu}-\frac12g_{\mu\nu}R=8\pi T_{\mu\nu},
\end{equation}
\begin{equation}
\label{Vlasoveq}
p^\alpha \partial_{x^\alpha}f -\Gamma^\alpha_{\beta\gamma}p^\beta p^\gamma \partial_{p^\alpha}
f=0,
\end{equation}
\begin{equation}
\label{Tdef}
T_{\alpha\beta}(x)=\int_{\pi^{-1}(x)} p_{\alpha}p_{\beta} f,
\end{equation}
where $p^\alpha$ define the momentum coordinates on the tangent bundle
conjugate to $x^\alpha$, where $\pi:P\to \mathcal{M}$ denotes the natural projection, and the integral 
in $(\ref{Tdef})$ is to be understood with respect to the natural
volume form on $\pi^{-1}(x)$.

We call $f$ the \emph{Vlasov field}.
The equation $(\ref{Vlasoveq})$ is just the statement in coordinates that
 $f$ be preserved
along geodesic flow on $P$.
 In physical language, $f$ describes thus the distribution
of so-called \emph{collisionless matter}, and sometimes we shall refer to solutions of
 the system $(\ref{Eeq})$--$(\ref{Tdef})$
as collisionless matter spacetimes.

For any null vector $V$, in view of the condition $f\ge0$, $(\ref{Tdef})$ and
$(\ref{Eeq})$, one obtains
the inequality
\begin{equation}
\label{dec}
R_{\mu\nu} V^\mu V^\nu \ge 0.
\end{equation}
Collisionless matter spacetimes thus satisfy	  the \emph{null convergence condition}.

For a full discussion of the Einstein-Vlasov system, see~\cite{ha:living, ar:vlasov}.

\section{$T^2$ symmetry}
We will say that a spacetime $(\mathcal{M},g)$ is $T^2$ symmetric  if the Lie 
group $T^2$ acts differentiably on $(\mathcal{M},g)$ by isometry, and the group orbits
are spatial.
See~\cite{berger} for a general discussion. 
The Lie algebra is spanned by two commuting Killing fields
$X$ and $Y$ which are nonvanishing.
We may normalise these so that the quantity
\begin{equation}
\label{Rdef}
R= g(X,X)g(Y,Y)-g(X,Y)^2
\end{equation}
gives the area element of the group orbits when multiplied by $X\wedge Y$. 
In particular, $X$ and $Y$ are nowhere vanishing in the spacetime.

The Gowdy case studied in~\cite{mw} is a special case of the above, when the
so-called twists of $X$ and $Y$ vanish. 
Note that in 
the vacuum case, if $(\mathcal{M},g)$ is globally hyperbolic and spatially compact, then either
it is Gowdy, or its spatial topology is $T^3$. 

\section{The main theorem}
\begin{theorem}
\label{MT}
Let $(\mathcal{M},g)$ be a globally hyperbolic $T^2$-symmetric spacetime with $C^2$ metric, 
with compact Cauchy
surface $\Sigma= T^3$ topologically,  
let $X$ and $Y$ be globally defined Killing fields spanning
the Lie algebra, let $R$ be as in $(\ref{Rdef})$.
Assume 
\begin{enumerate}
\item
All past incomplete causal geodesics $\gamma(t)$
satisfy $R(t)\to 0$, and
\item
$f: P\to \mathbb R$ is such that $g, f$ satisfy the Einstein-Vlasov system, 
with $f\in C^0$, and
\item
There exists a constant $\delta>0$ such that
for any open $\mathcal{U}\subset P\cap \pi^{-1}(\Sigma)$ we have that $f$ does not vanish identically
on $\mathcal{U}\cap \{p: g(p, X)^2+g(p,Y)^2<\delta\}$.
\end{enumerate}
Then $(\mathcal{M},g)$ is past inextendible as a $C^2$ Lorentzian manifold.
\end{theorem}

With the usual convention for time orientation,
Assumption 1 above has been shown in~\cite{mw} for
maximal developments of all sufficiently regular  $T^2$ initial
data sets (with topology $T^3$) for $(\ref{Eeq})$--$(\ref{Tdef})$,
for which $f$ does not vanish identically,
 provided that $f$ is initially compactly supported
in~$P\cap \pi^{-1}(\Sigma)$. 
As in~\cite{dr3}, this argument can easily be adapted to some class where the
data remain compactly supported in the tangential momentum directions, but
are allowed in other directions
to decay sufficiently fast initially with respect to linear coordinates on $P$. Assumption
2 of course holds for such maximal developments by definition.
Within this extended class, Assumption 3 can be viewed as a genericity assumption.
Thus the above theorem implies strong cosmic censorship (in its $C^2$-inextendibility
formulation) in the past direction.

On the other hand, the future inextendibility requirement for 
strong cosmic censorship holds for maximal developments
of arbitrary $T^2$-symmetric initial data (with topology $T^3$) for $(\ref{Eeq})$--$(\ref{Tdef})$
in view of the results of~\cite{treis, dr2}.
The results of~\cite{dr2} are more elementary than those here, but also rest on the extendibility of the Killing vectors.

Thus, Theorem~\ref{MT} implies strong cosmic censorship for $T^2$ symmetric
spacetimes with collisionless matter and spatial topology $T^3$.

\section{Cauchy horizon rigidity}
We show in this section that under the first assumption of Theorem~\ref{MT},
Cauchy horizons must inherit a certain rigidity, namely, at regular points,
the Ricci curvature in the direction of the null generator must vanish.

\begin{proposition}
\label{rigid}
Let $(\mathcal{M},g)$ be a globally hyperbolic $T^2$ symmetric spacetime as
in Theorem~\ref{MT}, satisfying Assumption 1, but
not necessarily Assumptions 2 and 3.
Suppose $(\mathcal{M},g)$ is past extendible and let $\mathcal{H}^-$ denote the
past Cauchy horizon of $\Sigma$ in the extension $(\tilde{\mathcal{M}},\tilde{g})$. 
Then there exists a dense subset $\tilde{S}\subset\mathcal{H}^-$, 
at which $T_p\mathcal{H}^-$ is a hyperplane whose orthogonal complement is spanned by
a null vector $V$, for which
\[
{\rm Ric}(V,V)\le 0.
\]
\end{proposition}

\begin{proof}
$\mathcal{H}^-$ is an achronal Lipschitz submanifold~\cite{he:lssst}.
By the results of~\cite{chrusciel}, 
it is differentiable on a dense subset $S$, on which its tangent
plane must clearly then be null.

In what follows we will relate the null generator to $T_p\mathcal{H}^-$
at regular points to the span of the Killing vector fields.

By the results of~\cite{dr2}, $X$ and $Y$ extend $C^2$ through $\mathcal{H}^-$.
(That is to say, they can be extended to $C^2$ vector fields on $\tilde{\mathcal{M}}$, 
not necessarily
Killing.) 

Let $p\in\mathcal{H}^-$ be regular. 
The vectors $X$ and $Y$ must clearly be tangent to $\mathcal{H}^-$ at $p$. This follows
since the integral curves of $X$, $Y$ through points of $\mathcal{M}$ stay in $\mathcal{M}$.

Define
\[
\mathcal{H}^-_2 = \{ X\wedge Y\ne 0\}\cap \mathcal{H}^-,
\]
and
\[
\mathcal{H}^-_1=\rm{int}( \{X\wedge Y=0, X\ne0, Y\ne0\}),
\]
where $\rm{int}$ denotes the interior with respect to the topology
of $\mathcal{H}^-$. 
These are clearly open subsets of $\mathcal{H}^-$.

\begin{lemma}
If $p\in \mathcal{H}^-_2$, then $X(p)$, $Y(p)$ span a null plane,
tangent to $\mathcal{H}^-$ if in addition $p\in S$. 
If $p\in \mathcal{H}^-_1$, then $X$ and $Y$ lie in
a null direction, again tangent to $\mathcal{H}^-$ if in addition $p\in S$. Finally,
\begin{equation}
\label{decomp}
\mathcal{H}^-=\overline{ \mathcal{H}^-_1\cup \mathcal{H}^-_2}.
\end{equation}
\end{lemma}
\begin{proof}
By the assumption that $R\to 0$ along any causal geodesic approaching $\mathcal{H}^-$,
it follows that $R$ extends to a $C^2$ function vanishing along $\mathcal{H}^-$.
From $(\ref{Rdef})$, it is clear that at points $p\in \mathcal{H}^-_2$,
the plane spanned by $X(p)$ and $Y(p)$ is null. The first statement of the Lemma 
follows in view also of the fact that $X$ and $Y$ are tangent to $\mathcal{H}^-$ at $S$.

To prove $(\ref{decomp})$, it is equivalent to prove that the set
\[
\{X=0\}\cup \{Y=0\} 
\]
has empty interior in $\mathcal{H}^-$.
This in turn will follow from the following statement: Let $Z$ be a vector field in the span
of $X$ and $Y$, such that $Z$ does not identically vanish on the spacetime. 
Since $R>0$ in the spacetime, in fact, it follows that $Z$ vanishes nowhere in the
spacetime.
Then $\{Z=0\}$ has empty interior in $\mathcal{H}^-$.

Let $Z$ be then  as above, and let $\mathcal{U}$ denote the interior of $\{Z=0\}$. 

We will first show that $\nabla Z$ vanishes identically in $\mathcal{U}$.
For $q\in \mathcal{U}\cap S$, 
let $E_1(q)$, $E_2(q)$, $L(q)$, $K(q)$ denote a null frame
where $E_1(q)$, $E_2(q)$, $K(q)$ are tangent 
to $\mathcal{H}^-$ at $q$.
For any vector $W(q)$, we compute:
\[
g(\nabla_{E_i}Z, W) = E_i g(Z,W)- g(Z,\nabla_{E_i}W)=0,
\]
\[
g(\nabla_KZ,W)=Kg(Z,W)-g(Z,\nabla_KW)=0,
\]
\[
g(\nabla_LZ,E_i)= -g(\nabla_{E_i}Z,L)=0,
\]
\[
g(\nabla_LZ,K)=-g(\nabla_KZ,L)=0,
\]
\[
g(\nabla_LZ,L)=0,
\]
where we have used the Killing property of $Z$ and its vanishing in $\mathcal{U}$. 
Thus $\nabla Z$ vanishes
identically in $\mathcal{U}\cap S$, 
and thus, by density and continuity, identically on $\mathcal{U}$. 

From the well known relation
\[
\nabla_\alpha\nabla_\beta Z_\gamma=R_{\alpha\beta\gamma\delta}Z^\delta
\]
which holds for any Killing vector field $Z$, by considering a family
of timelike geodesics transverse to $\mathcal{H}^-$, it follows
immediately that, if $\mathcal{U}\ne\emptyset$, then 
$Z$ must vanish identically in a neighborhood of $S\cap\mathcal{U}$
in $\mathcal{M}$. Since $Z$ does not vanish at any point of $\mathcal{M}$, 
and $S$ is dense in $\mathcal{U}$, we must have $\mathcal{U}=\emptyset$. 
This shows $(\ref{decomp})$.

We turn to show the second statement of the Lemma.
Consider a point $p \in \mathcal{H}^-_1\cap S$.
Completing $X$ to a $C^2$ frame $X$, $V_1$, $V_2$, $V_3$ for the tangent
bundle in a neighborhood of $p$, we may write
\[
Y=\alpha X+\beta_1V_1+\beta_2V_2+\beta_3V_3
\]
where $\beta_i$, $\alpha$ are $C^2$ functions. Since $\mathcal{H}^-_1$ is open,
we have
\begin{equation}
\label{theassum}
\beta_1=\beta_2=\beta_3=0
\end{equation}
in a neighborhood in $\mathcal{H}^-$ of $p$. 

From $[X,Y]=0$ we obtain
\[
(X\alpha) X+(X\beta_1) V_1+(X\beta_2 )V_2+(X\beta_3)V_3 =0,
\]
and thus $\alpha$ in particular is constant along integral curves of $X$.
On the other hand, for $V$ tangent to $\mathcal{H}^-$ at $p$, 
\begin{eqnarray*}
0	&=& g(\nabla_{V}Y, V) \\
 	&=& g(\nabla_{V}(\alpha X+\beta_1V_1+\beta_2V_2+\beta_3V_3),V) \\
	&=& \alpha g(\nabla_VX,V)+(V\alpha) g(X,V)+\sum\beta_i g(\nabla_VV_i,V)
				+\sum_i (V\beta_i)g(V_i,V)\\
	&=&(V\alpha)g(X,V).
\end{eqnarray*} 
If $X$ is spacelike, then the orthogonal space of $X$ is transverse to $\mathcal{H}^-$ at $p$,
and thus the above implies that $V\alpha=0$ for a dense set of directions, and thus by continuity
argument for all
directions tangent to $\mathcal{H}^-$. By an additional
continuity argument we obtain that $Y$ is a constant
multiple $aX$ in the connected component of $p$ in $\mathcal{H}^-_1$. 
Defining $Z= Y-a X$ and applying the previous result that $\{Z=0\}$ has empty
interior in $\mathcal{H}^-$, we obtain 
a contradiction. Thus $X$ is null.
\end{proof}

To proceed we will first need the following 
\begin{lemma}
\label{SG}
Let $p\in S$ and let $K$ denote a Killing vector
field such that $K(p)$ is null. Then
\begin{equation}
\label{gvwstne3}
\nabla_KK(p)=\kappa(p) K(p),
\end{equation}
for a real number $\kappa$.
\end{lemma}
\begin{proof}
Complete $K(p)$ to a null frame $K(p)$, $L(p)$, $E_1(p)$, $E_2(p)$ at $T_p\tilde{\mathcal{M}}$,
such that $E_1$, $E_2$ are in $T_p\mathcal{H}^-$.

Note first that 
\[
E_1g(K,K)(p)=E_2g(K,K)(p)=0,
\]
since $p$ is a local minimum of $g(K,K)$ along $\mathcal{H}^-$ to which both
$E_1$ and $E_2$ are tangent.
Thus, by the Killing equation, 
\[
0=E_1g(K,K)(p)=2 g(\nabla_{E_1}K,K)(p) = -g(\nabla_KK,E_1)(p),
\]
\[
0=E_2g(K,K)(p)=2 g(\nabla_{E_2}K,K)(p) = -g(\nabla_KK,E_2)(p).
\]
On the other hand $Kg(K,K)=0$ on account of the Killing equation,
and thus similarly we have $g(\nabla_K K,K)=0$. 
Thus, $\nabla_KK(p)$ is in the direction $K$, and $\kappa$ of $(\ref{gvwstne3})$
can be obtained
from
\[
\frac14Lg(K,K)(p) = \frac12 g(\nabla_LK,K)(p)=-\frac12 g(\nabla_KK,L)(p)\doteq \kappa.
\]
\end{proof}

We will need a further partition of $\mathcal{H}^-_1$ and $\mathcal{H}^-_2$. 
Applying Lemma~\ref{SG} to
points $p\in\mathcal{H}^-_1\cap S$ and the vector
$K=X$, we deduce $\nabla_XX=\kappa(p) X$.
By density and continuity, $\nabla_XX=\kappa X$ for all points of $\mathcal{H}^-_1$, where
$\kappa$ is a continuous function on $\mathcal{H}^-$. Define
\[
\mathcal{H}^-_{1, 0} =\{p \in\mathcal{H}^-_1, \kappa(p)= 0\}
\]
and
\[
\mathcal{H}^-_{1, reg} =\mathcal{H}^-_1 \setminus \mathcal{H}^-_{1, 0}.
\]
The set $\mathcal{H}^-_{1, reg}$ is clearly open.

On the other hand, let $p\in\mathcal{H}^-_2$, and
let $L$ denote a $C^2$ null vector field transverse to the $C^2$ distribution spanned
by $X$ and $Y$ in a a neighborhood of $p$. 
Consider the $C^1$ function $LR$ on $\tilde{\mathcal{M}}$.
$LR$ restricted to $\mathcal{H}^-$ is continuous. Define
\[
\mathcal{H}^-_{2,0}=\{p\in\mathcal{H}^-_2, LR=0\}
\]
and
\[
\mathcal{H}^-_{2, reg} =\mathcal{H}^-_2 \setminus \mathcal{H}^-_{2, 0}.
\]
Again, the set $\mathcal{H}^-_{2, reg}$ is clearly open. Note that $\mathcal{H}^-_{1,0}\cap S$
coincides with set of points $p$ where $\kappa(p)=0$ for any $\kappa(p)$
given by Lemma~\ref{SG}.

We have
\begin{eqnarray}
\label{teldec}
\nonumber
\mathcal{H}^-	&=&	\overline{\mathcal{H}^-_{1,reg}\cup \mathcal{H}^-_{2,reg}}
					\cup\mathcal{H}^-_{1,0}\cup\mathcal{H}^-_{2,0}\\
			&=&	\overline{\mathcal{H}^-_{1,reg}\cup \mathcal{H}^-_{2,reg}}
					\cup\overline{(\mathcal{H}^-_{1,0}\cup\mathcal{H}^-_{2,0})\cap S}.
\end{eqnarray}
%In fact, we have
%\begin{lemma}
%$\mathcal{H}^-_{1,0}$ is empty, i.e.
%\[
%\mathcal{H}^-=\overline{\mathcal{H}^-_{1,reg}\cup \mathcal{H}^-_{2}}.
%\]
%\end{lemma}
%\begin{proof}
%Let $p\in S\cap \mathcal{H}^-_{1,0}$. Let $X$ and $Y$ be as before. We have
%that $Y(p)=\alpha(p) X$. Set $a=\alpha(p)$, and consider the Killing vector
%field $Z=X-aY$. Clearly $Z(p)=0$. Moreover, $\nabla Z=0$. To see this,
%consider a null frame $X$, $L$, $E_1$, $E_2$ at $p$, where $E_1$ and $E_2$
%are unit vector fields tangential to $T_p\mathcal{H}^-$.  We may extend $E_1$ and $E_2$
%so as to be $C^2$ sections of the $C^2$ distribution orthogonal to $X$, say. 
%Recall that $Z$ is null
%in a neighborhood of $T_p$ and is in the direction of the null generator
%in $S\cap\mathcal{H}^-_{1,0}$ unless it vanishes. We compute
%\[
%g(\nabla_{E_i}Z,{E_j})(p)=E_ig(Z,E_j)(p)-g(Z,\nabla_{E_i}E_j)(p)=0.
%\]
%where we have used the fact that $g(Z,E_j)=0$ identically on $\mathcal{H}^-$ in a neighborhood
%of $p$, and $E_i$ is tangential to $\mathcal{H}^-$ at $p$.
%Similarly,
%\[
%g(\nabla_{E_i}Z,K)(p)=E_ig(Z,K)(p)-g(Z,\nabla_{E_i}K)(p)=0.
%\]
%On the other hand
%\[
%g(\nabla_{E_i}Z, L)(p)=E_i g(Z,L)-g(Z,\nabla_{E_i}L)
%\]
%AYTO DEN EINAI MHDEN
%\end{proof} 

We would like to do certain computations with frames ``adapted'' to the horizon $\mathcal{H}^-$. 
For this, we will need that the various parts of the horizon have sufficient regularity. Our first
result in this direction is the following
\begin{lemma}
\label{reglem}
$\mathcal{H}^-_{1,reg}$ is a $C^3$ hypersurface.
\end{lemma}
\begin{proof}
First we show that $\mathcal{H}^-_{1,reg}$ is $C^2$. Consider the function
\[
h=g(X,X).
\]
This extends $C^2$ through $\mathcal{H}^-$.
Clearly, if $\kappa\ne 0$, then $\nabla h\ne 0$, and thus, since then
$\mathcal{H}^-_{1,reg}\subset h^{-1}(0)$, we have then
that $\mathcal{H}^-_{1,reg}$ is $C^2$.

Now consider the $C^2$ orthogonal distribution to the one-dimensional distribution spanned
by $X$.  Since $\mathcal{H}^-_{1,reg}$ has been shown to be $C^1$ (in fact $C^2$), and its
normal coincides with $X$ on $S$, then its normal is in the direction of $X$ everywhere,
i.e.~its tangent space is the orthogonal complement of $X$.
Thus $\mathcal{H}^-_{1,reg}$ is an integral manifold of the above mentioned $C^2$
distribution, and, consequently, is in fact $C^3$.
\end{proof}

For $\mathcal{H}^-_{2,reg}$ we similarly show
\begin{lemma}
\label{1rl}
$\mathcal{H}^-_{2, reg}$ is a $C^3$ hypersurface.
\end{lemma}
\begin{proof}
As before, first we show that $\mathcal{H}^-_{2,reg}$ is $C^2$.
Recall that the function $R$ extends as a $C^2$ function through the boundary.
Moreover, 
\begin{equation}
\label{inviewof}
\mathcal{H}^-_{2,reg}\subset \{R=0\}.
\end{equation} 
Given $p\in\mathcal{H}^-_{2,reg}$,
since by definition $LR(p)\ne 0$ for some $L$ defined near $p$, 
it follows that $\nabla R\ne0$. Thus $\{R=0\}$ is a $C^2$
submanifold near $p$, which must thus coincide with the Lipschitz
manifold $\mathcal{H}^-_{2,reg}$ in view of $(\ref{inviewof})$.
Thus, $\mathcal{H}^-_{2,reg}$ is $C^2$.

To show additional regularity, we shall construct a $3$-dimensional $C^2$
distribution tangent to $\mathcal{H}^-_{2,reg}$.

The vectors $X$ and $Y$ span a two-dimensional distribution
satisfying $XR=0$, $YR=0$. Let $E$ be a $C^2$ section
of the $C^2$ distribution orthogonal to that spanned by $X$ and $Y$. 
Consider a regular point $p \in \mathcal{H}^-_2\cap S$.
We may choose Killing fields $\tilde{X}$, $\tilde{Y}$
so that $\tilde{X}(p)$ is null, normalised so that again
$R=g(\tilde{X},\tilde{X})g(\tilde{Y},\tilde{Y})-g(\tilde{X},\tilde{Y})^2$.
We have at $p$
\begin{eqnarray*}
ER(q)	&=&	Eg(\tilde{X},\tilde{X})g(\tilde{Y},\tilde{Y})+g(\tilde{X},\tilde{X})Eg(\tilde{Y},\tilde{Y})
				-2g(\tilde{X},\tilde{Y})Eg(\tilde{X},\tilde{Y})\\
		&=&	g(\nabla_{\tilde{X}}\tilde{X},E) g(\tilde{Y},\tilde{Y})\\
		&=& g(\tilde\kappa \tilde{X},E) g(\tilde{Y},\tilde{Y})\\
		&=& 0,
\end{eqnarray*}
where we have used $g(\tilde{X},\tilde{X})=0$, $g(\tilde{X},E)=0$ and Lemma~\ref{SG}. Thus, 
$ER=0$ on $\mathcal{H}^-_2\cap S$
and thus by continuity and density, on $\mathcal{H}^-_2$.

One easily sees that $\mathcal{H}^-_2$ is an integral manifold of the distribution
spanned by the $C^2$ vector fields $E$, $X$, and $Y$, and is thus $C^3$.
(We in fact only need the $C^2$ statement in what follows.)
\end{proof}

\begin{lemma}
\label{difficult}
For $q\in \mathcal{H}^-_{2,reg}$, let $K$ denote a Killing vector field
such that $K(q)$ is a null generator for $T_q\mathcal{H}^-_{2,reg}$. 
$K$ can be completed to a $C^2$ frame $K$, $L$, $E_1$, $E_2$ 
for the tangent bundle of $\tilde{\mathcal{M}}$ near $q$,
such that
at $q$, the vectors $K(q), N(q), E_1(q), E_2(q)$ constitute
a null frame, the vector field $E_1$ is Killing in $\mathcal{M}$, $E_2$ is
tangent to the Cauchy horizon and
\begin{equation}
\label{evallakt}
g(K, \nabla_{E_1} E_1)(q)=0,
\end{equation}
\begin{equation}
\label{yeniyIldIz*evallakt}
E_1E_1g(K,K)(q)=0,
\end{equation}
\begin{equation}
\label{yeniyildiz**1}
g(K, \nabla_{E_2}E_2 ) (q)=0,
\end{equation}
\begin{equation}
\label{yeniyildiz**2}
E_2E_2 g(K,K)(q)= 8(g(\nabla_{E_2}K,E_1))^2(q).
\end{equation}
\end{lemma}
\begin{proof}
Let $E_1$ denote a Killing field such that $g(E_1,E_1)(q)=1$, and let $E_2$
denote (as in the proof of Lemma~\ref{1rl}) 
a $C^2$ section (in a neighborhood of $q$) of the distribution orthogonal to 
that spanned by the Killing fields, with the additional restriction that $g(E_2,E_2)(q)=1$.

To see $(\ref{yeniyIldIz*evallakt})$, note first
that  $Y g(X,X)=0$ everywhere in $\mathcal{M}$,
and in addition, certainly $Xg(X,X)=0$. Similarly
$X g(Y,Y)=0$, $Y g(Y,Y)=0$. Thus $K g(E_1,E_1)=0$
identically, in particular $(\ref{evallakt})$ holds, and
$E_1 E_1 g(K,K)=0$ identically, in particular $(\ref{yeniyIldIz*evallakt})$.

For $(\ref{yeniyildiz**1})$, just note
\[
g(K,\nabla_{E_2}E_2)=E_2g(K,E_2)-g(\nabla_{E_2}K,E_2)=E_2g(K,E_2)=0,
\]
where we have used that $g(K,E_2)=0$ identically, as well as the Killing property
of $K$.

Note also that $E_2 g(K,K)(q)=0$ since $q$ is a local minimum of $g(K,K)$ restricted
to $\mathcal{H}^-$. Now, by the regularity of Lemma~\ref{1rl}, it follows in particular, that the integral
curves of $E_2$ through points of $\mathcal{H}^-_{2,reg}$ stay on $\mathcal{H}^-_{2,reg}$ for short
time,
and thus, since 
\[
g(K,K)g(E_1,E_1)-g(K,E_1)^2=0
\]
identically on $\mathcal{H}^-$, we may differentiate twice at $q$ in the direction $E_2$
\begin{align*}
&E_2E_2 g(K,K) g(E_1,E_1)(q)
+2E_2 g(K,K) E_2 g(E_1,E_1)(q)\\
&+
g(K,K) E_2E_2g(E_1,E_1)(q) \\
&- 2(E_2 g(K,E_1))^2(q)-2g(K,E_1) E_2E_2 g(K,E_1)=0
\end{align*}
to obtain
\begin{equation}
\label{firsttry}
E_2E_2 g(K,K)(q)= 2(E_2 g(K,E_1))^2.
\end{equation}

On the other hand,
\begin{eqnarray*}
E_2g(K,E_1)	&=&	g(\nabla_{E_2}K,E_1)+g(K,\nabla_{E_2}E_1)\\
			&=&	g(\nabla_{E_2}K,E_1)-g(E_2,\nabla_KE_1)\\
			&=&	g(\nabla_{E_2}K,E_1)-g(E_2,\nabla_{E_1}K)\\
			&=&	g(\nabla_{E_2}K,E_1)+g(E_1,\nabla_{E_2}K)\\
			&=&	2g(\nabla_{E_2}K,E_1),
\end{eqnarray*}
where we have used the Killing property of $E_1$ and $K$, as well as $[E_1,K]=0$.
The above and $(\ref{firsttry})$ gives $(\ref{yeniyildiz**2})$.
\end{proof}

%\noindent{\bf Remark.} The construction of the vector field $E_2$ can easily
%be seen to show that $\mathcal{H}^-_2$ is in fact a $C^3$ hypersurface,
%as it is an integral manifold of the $C^2$ distribution spanned by
%$K$, $E_1$ and $E_2$. We shall not need this fact here.
%\vskip1pc
%In the case of $\mathcal{H}^-_{1,reg}$, we will need a frame adapted to the hypersurface.
%For this, we will need to know some regularity. We have

\begin{lemma}
\label{kappane0case}
Let $q\in \mathcal{H}^-_{1,reg}$, let $K$ denote the Killing vector field $X$.
Then $K$
can be completed to a $C^2$ frame $K$, $L$, $E_1$, $E_2$ 
for the tangent bundle of $\tilde{\mathcal{M}}$ near $q$,
such that
at $q$, the vectors $K(q)$, $L(q)$, $E_1(q)$, $E_2(q)$ constitute
a null frame, and the relations $(\ref{evallakt})$, $(\ref{yeniyIldIz*evallakt})$,
 $(\ref{yeniyildiz**1})$ hold, as well as $(\ref{yeniyildiz**2})$, where
now both sides of the equality vanish. 
\end{lemma}
\begin{proof}
Choose unit vectors $E_1(q)$ and $E_2(q)$ tangent to $\mathcal{H}^-$ at $q$
and extend these as $C^2$ vector fields in $\tilde{\mathcal{M}}$ tangent to 
$\mathcal{H}^-$ in a neighborhood of $q$. This is possible in view of Lemma~\ref{reglem}.
Define $L$ to be an arbitary extension of a null vector $L(q)$ orthogonal to  $E_1(q)$
and $E_2(q)$, such that $g(L(q),K(q))=-2$.

Recall that $g(K,K)=0$ identically on $\mathcal{H}^-_{1,reg}$. Since the integral curves
of the $E_i$ remain on $\mathcal{H}^-_{1,reg}$ for short time, we have 
$(\ref{yeniyIldIz*evallakt})$
and $E_2E_2g(K,K)=0$.
Again, since  the $E_i$ are tangent to $\mathcal{H}^-_{1,reg}$ in a neighborhood of $q$, 
we have that $g(K,E_i)=0$ on this neighborhood, and thus in particular
\[
g(K,\nabla_{E_i}E_i)(q) = E_i g(K, E_i)(q)-g(\nabla_{E_i}K,E_i)(q)=E_ig(K,E_i)(q)=0
\]
where we have also used the Killing property of $K$. This gives $(\ref{evallakt})$
and, in view also of $E_2E_2g(K,K)=0$ derived earlier, $(\ref{yeniyildiz**2})$,
with both sides equal to $0$.
\end{proof}

For the case of $(\mathcal{H}^-_{1,0}\cup \mathcal{H}^-_{2,0})\cap S$, 
it is not clear that one can obtain the regularity necessary
to consider adapted $C^2$ (or even $C^1$) frames as above. The highly degenerate nature of this case, however,
and the nature of the arguments that follow mean that the following result will suffice for us:
\begin{lemma}
\label{kappa=0case}
Let $q\in(\mathcal{H}^-_{1,0}\cup\mathcal{H}^-_{2,0})\cap S$, 
and let $K$ denote a Killing vector field such that $K(p)$ is null, and
consider any $C^2$ frame $K$, $L$, $E_1$, $E_2$  for the tangent bundle near $q$, extending
$K$, where $K(q)$, $L(q)$, $E_1(q)$, $E_2(q)$ is a null frame such that $E_1(q)$ and $E_2(q)$
are tangent to $\mathcal{H}^-$ at $q$. Then
\[
E_1E_1g(K,K)(q)\ge 0,
\]
\[
E_2 E_2 g(K,K)(q)\ge 0.
\]
\end{lemma}
\begin{proof}
Let $K$, $L$, $E_1$, $E_2$ be any frame as in the statement of the Lemma. (There
are no obstructions to the construction of such a frame. In the case of $q\in\mathcal{H}^-_{1,0}$,
we may take $K=X$.)

In the case of $q\in\mathcal{H}^-_{1,0}\cap S$ recall $\kappa$ defined by $\nabla_XX=\kappa X$.
For $q\in\mathcal{H}^-_{2,0}\cap S$, consider the $\kappa$ of Lemma~\ref{SG} applied to $K$.
Consider an arbitrary vector field $W$ transverse to $\mathcal{H}^-$ at $q$. 
By the condition $\kappa(q)=0$, we have that
$W g(K,K)(q)=0$. On the other hand, we have that $g(K,K)>0$ in the spacetime.
Thus $W(W g(K,K) )(q)\ge 0$. Since any $E_i$ is a limit of transversal vectors, it follows
that $E_iE_i g(K,K)(q)\ge 0$, as desired.
\end{proof}

We may now complete the proof of Proposition~\ref{rigid}.
Let 
\[
q\in\mathcal{H}^-_{1,reg}\cup\mathcal{H}^-_{2,reg}
\cup(\mathcal{H}^-_{1,0}\cup \mathcal{H}^-_{2,0}\cap S),
\]
and let $K$, $L$, $E_1$, $E_2$ be a frame as in Lemma~\ref{difficult}, Lemma~\ref{kappane0case},
or Lemma~\ref{kappa=0case}.
Recall now the identity
\begin{equation}
\label{theid}
\Box g(K,K)=-2{\rm Ric} (K,K) +  2 g(\nabla K,\nabla K).
\end{equation}

We evaluate $(\ref{theid})$ at $q$. In view of the properties of the frame we obtain
\begin{eqnarray*}
\Box g(K,K)(q) &=&
-2\nabla^2_{L,K}g(K,K)
+\nabla^2_{E_1,E_1}g(K,K)+\nabla^2_{E_2,E_2}g(K,K)\\
&=&-2L (K(g(K,K))+2\nabla_{{\nabla_L}K}g(K,K)
+E_1(E_1(g(K,K)))\\
&&\hbox{} + E_2(E_2(g(K,K)))
-\nabla_{\nabla_{E_1}E_1}g(K,K)-\nabla_{\nabla_{E_2}E_2}g(K,K)\\
&=&E_1(E_1(g(K,K)))+E_2(E_2(g(K,K)))+
4g(\nabla_{\nabla_L K}K,K)\\
&&\hbox{}-2g(\nabla_{\nabla_{E_1}E_1}K,K)
-2g(\nabla_{\nabla_{E_2}E_2}K,K)\\
&=&E_1(E_1(g(K,K)))+E_2(E_2(g(K,K)))-
4g(\nabla_K K,\nabla_L K)\\
&&\hbox{}+2g(\nabla_KK,\nabla_{E_1}E_1)
+2g(\nabla_KK,\nabla_{E_2}E_2)\\
&=&E_1(E_1(g(K,K)))+E_2(E_2(g(K,K)))-
8\kappa^2\\
&&\hbox{} +2g(\nabla_KK,\nabla_{E_1}E_1)
+2g(\nabla_KK,\nabla_{E_2}E_2)\\
&=&E_1(E_1(g(K,K)))+E_2(E_2(g(K,K)))-
8\kappa^2\\
&&\hbox{} +2\kappa g(K,\nabla_{E_1}E_1)
+2\kappa g(K,\nabla_{E_2}E_2).
\end{eqnarray*}

In the case of $q\in\mathcal{H}^-_{2,reg}$, from Lemma~\ref{difficult} we obtain
\begin{equation}
\label{thiscase}
\Box g(K,K)(q)=-8\kappa^2+8(g(\nabla_{E_2}K,E_1))^2(q).
\end{equation}
In the case of $q\in \mathcal{H}^-_{1,reg}$, from Lemma~\ref{kappane0case} we obtain
\begin{equation}
\label{thatcase}
\Box g(K,K)(q)=-8\kappa^2.
\end{equation}
Finally, in the case of $q\in(\mathcal{H}^-_{1,0}\cup \mathcal{H}^-_{2,0})\cap S$, from
Lemma~\ref{kappa=0case} we obtain
\begin{equation}
\label{3rdcase}
\Box g(K,K)(q) \ge 0.
\end{equation}
On the other hand,
\begin{eqnarray*}
2 g(\nabla K, \nabla K)(q)	 &=&-4 g(\nabla_L K,\nabla_K K)+ 2 g(\nabla_{E_1}K,\nabla_{E_1}K)
		+2 g(\nabla_{E_2}K,\nabla_{E_2}K)\\
&=&		-8\kappa^2+ 2(g(\nabla_{E_1}K,E_1))^2+2(g(\nabla_{E_1}K,E_2))^2\\
&&\hbox{}
		-4g(\nabla_{E_1}K,K)g(\nabla_{E_1}K,L)+2(g(\nabla_{E_2}K,E_2))^2 \\
&&	\hbox{}+2(g(\nabla_{E_2}K,E_1))^2
				-4g(\nabla_{E_2}K,K)g(\nabla_{E_2}K,L) \\
&=&-8\kappa^2 +4 (g(\nabla_{E_2} K,E_1))^2.
\end{eqnarray*}
We thus have
\begin{equation}
\label{not0}
{\rm Ric} (K, K)= -2 (g(\nabla_{E_2} K,E_1))^2
\end{equation}
in $\mathcal{H}^-_{2,reg}$,
\begin{equation}
\label{heusleralso}
{\rm Ric}(K,K)=0
\end{equation}
in $\mathcal{H}^-_{1,reg}$, and
\[
{\rm Ric}(K,K)\le 0
\]
in $(\mathcal{H}^-_{1,0}\cup\mathcal{H}^-_{2,0})\cap \mathcal{S}$.
In all cases,
\[
{\rm Ric}(K,K)\le 0.
\]
for any $q\in\tilde{S}\doteq((\mathcal{H}^-_{1,0}\cup\mathcal{H}^-_{2,0})\cap S)\cup\mathcal{H}^-_{1,reg}
\cup\mathcal{H}^-_{2,reg}$. 
By $(\ref{teldec})$, this suffices to obtain the Proposition.
\end{proof}

In the Gowdy case, the right hand side of $(\ref{not0})$ vanishes in view of the twist-free
condition for the Killing fields. More on this in Section~\ref{aside}.

\begin{proposition}
\label{=0}
Let $(\mathcal{M},g)$ be as in the statement of Proposition~\ref{rigid}, and assume
that it satisfies in addition 
the null convergence condition~$(\ref{dec})$.
Then 
\begin{equation}
\label{mndevizetai}
{\rm Ric}(K,K)(q)=0,
\end{equation}
on $\tilde{S}$.
\end{proposition}
\begin{proof}
This follows from Proposition~\ref{rigid} by
a simple continuity argument. Extend $K$ in a neighborhood of $q$ 
to be a $C^2$ null vector. Were ${\rm Ric}(K,K)(q)<0$, this would have to hold for
a point $p\in\mathcal{M}$, and this contradicts $(\ref{dec})$.
\end{proof}
The above Proposition clearly applies to $(\mathcal{M},g)$ satisfying Assumptions 1 and 2 
of Theorem~\ref{MT}, since collisionless matter spacetimes satisfy~$(\ref{dec})$.

\section{Aside: Killing horizons}
\label{aside}
Recall that we call a $C^1$ hypersurface $\mathcal{H}$ a \emph{Killing horizon} if its normal bundle is spanned by a null vector field which is Killing on $\mathcal{H}$.

For a $C^2$ Killing horizon with $K$ a null Killing vector field spanning the normal bundle, 
Proposition 6.15 of Heusler~\cite{unique} implies ${\rm Ric}(K,K)=0$.\footnote{The regularity
assumptions are not made precise in this proposition, but $C^2$ is certainly sufficient.}
Thus, since 
$\mathcal{H}^-_{1,reg}$ is a $C^3$ Killing horizon, the result 
$(\ref{heusleralso})$ could alternatively have been deduced from this proposition,
whose proof in any case is essentially the computation we have done above.

In fact we have
\begin{proposition}
Let $(\mathcal{M},g)$ be as in the statement of Proposition~\ref{rigid}, and assume
that it is in fact Gowdy symmetric. Then $\mathcal{H}^-_{1,reg}\cup \mathcal{H}^-_2$
is a $C^3$ Killing horizon. 
\end{proposition}

\begin{proof}
For $\mathcal{H}^-_{1,reg}$, there is nothing to say, in view of the previous. It suffices
thus to show that $\mathcal{H}^-_2$ is a $C^3$ Killing horizon.

The arguments below are also inspired
by a computation in~\cite{unique}. The result for $\mathcal{H}^-_{2,reg}$
could alternatively be deduced from the results of~\cite{carter}. Because
$\mathcal{H}^-_2$ is \emph{a priori} only Lipschitz, 
it is not clear that the results of~\cite{unique, carter}
 can be applied directly, and we thus give here a self-contained argument.

Without loss of generality, assume $X(p)=K(p)$ is null for some $p\in\mathcal{H}^-_{2}$,
and consider the vector field $\hat{K}$ defined by
\begin{equation}
\label{definedby}
\hat{K}= X- g(X,Y)(g(Y,Y))^{-1}Y.
\end{equation}
This is clearly null in a neighborhood of $p$ on $\mathcal{H}^-_{2}$, and is orthogonal
to $Y$ in a neighborhood in $\mathcal{M}$ of $p$. Let $E_2$ be a $C^2$ section of the
$C^2$ distribution orthogonal to that spanned by $X$ and $Y$ near $p$.
We will show that
\begin{equation}
\label{gvwsto}
E_2 (g(X,Y)(g(Y,Y))^{-1}) = 0.
\end{equation}
We compute
\begin{eqnarray*}
E_2 (g(X,Y)(g(Y,Y))^{-1}) &=&g(Y,Y)^{-2}\big(2g(Y,Y) g(\nabla_{E_2}X,Y)\\
					&&\hbox{}-2g(\nabla_{E_2}Y,Y)g(X,Y)\big)\\
		&=&g(Y,Y)^{-2}\big(-2g(Y,Y) g(\nabla_{Y}X,E_2)\\
			&&\hbox{}+2g(\nabla_{Y}Y,E_2)g(X,Y)\big)\\
		&=&g(Y,Y)^{-2}\big(-2g(Y,Y) g(\nabla_{X}Y,E_2)\\
			&&\hbox{}+2g(\nabla_{Y}Y,E_2)g(X,Y)\big)\\
		&=&-2g(Y,Y)^{-1}\big( g(\nabla_{X}Y,E_2)\\
			&&\hbox{}-g(X,Y)g(Y,Y)^{-1}g(\nabla_{Y}Y,E_2)\big)\\
		&=&-2g(Y,Y)^{-1} g(\nabla_{X-g(X,Y)g(Y,Y)^{-1}Y}Y,E_2)\\
		&=&0.
\end{eqnarray*}
In the above, we have used $[X,Y]=0$, the Killing property and the twist 
free property of the Killing fields.

Clearly, an identity analogous to $(\ref{gvwsto})$ holds when $E_2$ is replaced
by $X$ or $Y$.

Thus, let $W$ be any vector field in the span of $E_2$, $X$ and $Y$.
Let $p_i\to p$ be a sequence where $p_i\in\mathcal{M}$, the original spacetime.
It is clear that the integral curves of $W$ through $p_i$ cannot meet
$\mathcal{H}^-_{2}$ in a small neighborhood of $p$ for small time. 
For suppose $\gamma_i$ was such a curve. There would be a first point
of $\gamma_i$ meeting $\mathcal{H}^-_2$, i.e., there would be 
a segment $\gamma_i([t_0,t_1])$ with $\gamma_i([t_0,t_1))\subset\mathcal{M}$,
$\gamma_i(t_0)=p_i$,
and $\gamma_i(t_1)\in\mathcal{H}^-_2$. 
The above computation shows that $W(g(X,Y)/g(Y,Y))=0$, identically on $\gamma([t_0,t_1])$. 
Moreover, this remains true
 if $X$, and $Y$ are replaced by any other Killing vector fields $\tilde{Y}\ne \tilde{X}$,
 $g(\tilde{Y},\tilde{Y})>0$.
Choose $\tilde{X}$ such that $\tilde{X}$ is null at the point
$\gamma(t_1)\in\mathcal{H}^-_2$. We have
 $g(\tilde{X},\tilde{Y})(\gamma(t_1))=0$, and thus $g(\tilde{X},\tilde{Y})=0$ identically
 along $\gamma([t_0,t_1])$.
But at $p_i$, $\tilde{X}$ and $\tilde{Y}$ 
span a spacelike two-plane, thus there is a unique direction on this
plane
orthogonal to $\tilde{X}$ at $p_i$. Applying the above with a Killing field $\tilde{Y}\ne \tilde{X}$
 such that $\tilde{X}$ and $\tilde{Y}$ are \emph{not} orthogonal at $p_i$, we obtain a contradiction.

Since integral curves of $W$ through $p$ are limits of integral curves of $W$ through $p_i$,
it follows that these curves must remain on the boundary of $\mathcal{M}$, i.e.~on
$\mathcal{H}^-_{2}$.
One thus easily sees
that $\mathcal{H}^-_{2}$ is locally an integral manifold of the $C^2$ distribution
spanned by $X$, $Y$, and $E_2$, and thus $C^3$.

The above now shows that $g(X,Y)=0$ along $\mathcal{H}^-_{2}$ for any choice of
$Y$, and thus $X$ is null in the connected component of $p$. We have thus obtained
 that $\mathcal{H}^-_2$ is a $C^3$ Killing horizon, as desired.
\end{proof}

It would be nice to obtain that $\mathcal{H}^-_1$ is $C^3$. This would show that there is
a dense open subset of $\mathcal{H}^-$ which is a $C^3$ Killing horizon.
See also~\cite{cl}.

Finally, we also note the following
\begin{proposition}
\label{kh}
Let $(\mathcal{M},g)$ be as in the statement of Proposition~\ref{rigid},
and suppose that $\mathcal{M}$ is vacuum, i.e.~${\rm Ric}=0$ identically. Then
$\mathcal{H}^-_{1,reg}\cup \mathcal{H}^-_{2,reg}$ is a $C^3$ Killing horizon.
\end{proposition}
\begin{proof}
For $\mathcal{H}^-_{1,reg}$ there is nothing to show.
For $\mathcal{H}^-_{2,reg}$ in the Gowdy case, there is nothing to show in view of the previous proposition.

Let us assume thus  that the spacetime is not Gowdy, and let $p\in\mathcal{H}^-_{2,reg}$. Let
$K$ be a Killing vector field such that $K(p)$ is null. Formulas
$(\ref{mndevizetai})$ and $(\ref{not0})$ together imply that $g(\nabla_{E_2}K,E_1)=0$.
This implies that the twist quantity of $K$ vanishes. In the vacuum case, this implies
that this quantity vanishes on all of $\mathcal{M}$. Since there is a \emph{unique} Killing
vector field whose twist constant vanishes (in view of the assumption that the spacetime
$\mathcal{M}$ is \emph{not} Gowdy) it follows that the null Killing vector field at any other $q\in
\mathcal{H}^-_{2,reg}$ must be $K(q)$. Thus, $\mathcal{H}^-_{2,reg}$ is a Killing horizon.
\end{proof}

\section{The contradiction}
For the following proposition, the reader should again compare with~\cite{dr3}. The considerations
in the present paper are again easier, in view of 
the fact that the Killing fields are globally defined and do not
vanish in the original spacetime.
\begin{proposition}
\label{contra}
Let $(\mathcal{M},g)$ be as in the statement of Theorem~\ref{MT}. Then for a dense subset
of the dense subset
$\tilde{S}$ of Proposition~\ref{rigid},
\begin{equation}
\label{tosatisfy}
{\rm Ric}(K,K)>0.
\end{equation}
\end{proposition}
\begin{proof}
We will show that given any $p\in S$, where $S$ is as in the proof of Proposition~\ref{rigid},
and any open neighborhood $\mathcal{V}\subset\mathcal{H}^-$ of
$p$ in $\mathcal{H}^-$, there exists a $q\in \mathcal{V}\cap\tilde{S}$ satisfying ${\rm Ric}(K,K)>0$.

Let $K(p)$, $L(p)$, $E_1(p)$, $E_2(p)$ be null frame at $p$ such that $X$ and $Y$ lie
in the plane spanned by $K(p)$ and $E_1(p)$. Let $X(p)=aK(p)+bE_1(p)$, $Y(p)=cK(p)+dE_1(p)$. 
We have that $\max (|a|,|c|)>0$.
Consider the geodesic through $p$ with 
\[
\dot\gamma(p)= \frac14\delta (\max(|a|,|c|))^{-1}L
+ 2\delta^{-1}(\max(|a|,|c|)K.
\]
We clearly have $g(\dot\gamma,\dot\gamma)=-1$, while
\begin{equation}
\label{ine1}
|g(\dot\gamma, X)(p)| = \left|\frac14\delta g(L,K)a (\max(|a|,|c|))^{-1}\right| \le \delta/2,
\end{equation}
\begin{equation}
\label{ine2}
|g(\dot\gamma,Y)(p)| = \left|\frac14\delta g(L,K)c (\max(|a|,|c|))^{-1}\right| \le \delta/2.
\end{equation}
Since $\gamma$ is a geodesic and $X$, and $Y$
are Killing we have $\dot\gamma g(\dot\gamma,X)=0$, $\dot\gamma g(\dot\gamma,Y)=0$
and thus the inequalities $(\ref{ine1})$, $(\ref{ine2})$ 
hold throughout $\gamma$ in $\mathcal{M}$.

Now, $\gamma$ must intersect the Cauchy hypersurface at some time $T$.
By continuity of geodesic flow, for every neighborhood $\mathcal{V}$ in $\mathcal{H}^-$ of $p$  
there exists a
neighborhood $\mathcal{U}_0$ of $\gamma'(T)\in P$ in the topology of $P\cap \pi^{-1}(\Sigma)$
such that geodesics with initial condition on $\mathcal{U}$ intersect $\mathcal{H}^-$
transversally in $\mathcal{V}$. So select $\mathcal{U}_0$ so that this is the case for
our chosen $\mathcal{V}$, and then, in view of Assumption 2 and the continuity of $f$,
select $\mathcal{U}_1$ such that $f>\epsilon>0$ on an open $\mathcal{U}_1\subset \mathcal{U}_0$.
By the properties of the geodesic flow, the projection on $\mathcal{H}^-$ of the set
of all geodesics with initial condition in $\mathcal{U}_1$ contains a nonempty
open set $\mathcal{V}_1$. 

Let $q\in \tilde{S}\cap \mathcal{V}_1\ne\emptyset$ and let $V$ be a null generator for $\mathcal{H}^-$ at $q$.
By Proposition~\ref{rigid}, ${\rm Ric}(V,V)=0$. On the other hand, since $f$ can easily be
seen to extend continuously
to $P\cap\pi^{-1}(\mathcal{H}^-)$, we have that $f>0$ at some point of 
$P\cap\pi^{-1}(q)$, and thus in an open set. In particular, the integral defined by
$(\ref{Tdef})$ is strictly positive at $q$ when contracted twice with the null vector $V$.

Extend $V$ arbitrarily as a $C^2$ null vector field in a neighborhood of $q$.
In the original spacetime $\mathcal{M}$, the right hand side of $(\ref{Tdef})$
is equal to ${\rm Ric}(V,V)$ in view of $(\ref{Eeq})$. Taking a sequence of
points $q_i\to q$, with $q_i\in \mathcal{M}$, then ${\rm Ric}(V,V)(q_i)\to
{\rm Ric}(V,V)(q)$ by the fact that $\tilde{\mathcal{M}}$ is assumed $C^2$. 
By Fatou's lemma, the 
right hand side of $(\ref{Tdef})$ contracted twice with $V$ at $q$
is less than or equal to the limit of its value at $q_i$ when contracted twice
with $V$. The former
we have just shown to be strictly positive, while the latter equals ${\rm Ric}(V,V)(q)$. 
Thus ${\rm Ric} (V,V)(q)>0$, 
as desired.
\end{proof}

The proof of Theorem~\ref{MT} follows immediately from Propositions~\ref{=0}
 and~\ref{contra}.

\section{Comments}
The techniques of this paper are tied heavily to continuity of the curvature tensor.
The paper thus does not address weaker conditions of inextendibility, for instance,
inextendibility as a $C^0$ metric, which
may be more correct from a physical point of view. See~\cite{chr:givp, md:si, md:cbh}.

It is clear that Assumption 2 of Theorem~\ref{MT} can be weakened to the assumption
that the equations $(\ref{Eeq})$--$(\ref{Tdef})$ are satisfied with $(\ref{Tdef})$ replaced by
\[
T_{\alpha\beta}(x)=\tilde{T}_{\alpha\beta}(x)+ \int_{\pi^{-1}(x)} p_{\alpha}p_{\beta} f
\]
where $\tilde{T}_{\alpha\beta}(x)$ satisfies $\tilde{T}_{\alpha\beta}V^\alpha V^\beta\ge 0$
for all null $V^\alpha$. Thus strong cosmic censorship can be shown in $T^2$ symmetry
when the Einstein-Vlasov
system is extended to include other matter fields, \emph{provided that Assumption 1 can
also be shown for the maximal development of suitable data}.

\section{Acknowledgements}
M.D.~thanks the Albert Einstein Institute and A.D.R.~thanks the 
University of Cambridge for hospitality during visits. Some of this research was carried
out during the progamme ``Global Problems in Mathematical Relativity'' at the Isaac Newton
Institute of the University of Cambridge. M.D.~is supported in part by NSF grant
DMS-0302748 and a Marie Curie International Re-integration grant from the European 
Commission.

\end{document}